# MACU-Net for Semantic Segmentation of Fine-Resolution Remotely Sensed Images

Rui Li*, Chenxi Duan*, Shunyi Zheng, Ce Zhang and Peter M. Atkinson

*Abstract*—Semantic segmentation of remotely sensed images plays an important role in land resource management, yield estimation, and economic assessment. U-Net, a deep encoder-decoder architecture, has been used frequently for image segmentation with high accuracy. In this Letter, we incorporate multi-scale features generated by different layers of U-Net and design a multi-scale skip connected and asymmetric-convolution-based U-Net (MACU-Net), for segmentation using fine-resolution remotely sensed images. Our design has the following advantages: (1) The multi-scale skip connections combine and realign semantic features contained in both low-level and high-level feature maps; (2) the asymmetric convolution block strengthens the feature representation and feature extraction capability of a standard convolution layer. Experiments conducted on two remotely sensed datasets captured by different satellite sensors demonstrate that the proposed MACU-Net transcends the U-Net, U-NetPPL, U-Net 3+, amongst other benchmark approaches. Code is available at https://github.com/lironui/MACU-Net.

*Index Terms*—fine-resolution remotely sensed images, asymmetric convolution block, semantic segmentation

## I. Introduction

Semantic segmentation using remotely sensed images plays a critical role in a wide range of applications, such as land resource management, yield estimation, and economic assessment [1-3]. As an evolving technology, various classifiers have been developed for semantic segmentation in the remote sensing community, including orthodox methods (e.g. distance-based measures) [4], and machine learning techniques such as the support vector machine [5] and random forest [6]. However, the high dependency on hand-crafted features or mid-level semantic characteristics restricts the flexibility and adaptability of these traditional approaches [7].

Recently, Convolutional Neural Networks (CNN) [8] have demonstrated their strong capacity to capture nonlinear and hierarchical feature representations in an automatic fashion, influencing the field of computer vision (CV) significantly [9]. For semantic segmentation, encoder-decoder frameworks such as SegNet [10] and U-Net [11] have become commonly used schemes. Typically, feature maps generated by the encoder comprise low-level and fine-grained information, whereas those generated by the decoder contain high-level and coarse-grained semantic information [12]. As a bridge between low-level and high-level feature maps, skip connections are extra connections between encoders and decoders in the network, which are employed to boost the ability for semantic extraction within the encoder-decoder framework.

Despite its elegant structure and various achievements, insufficient utilization of information flow is a bottleneck that impedes the potential of the raw U-Net architecture. To address this issue, in U-Net++ [12], plain skip connections are substituted by nested and dense skip connections, which enhance the power of the skip connections and narrow the semantic gap between the encoder and decoder. However, neither the raw U-Net nor the U-Net++ extracts fully the multi-scale features within the network. Thus, full-scale skip connections are designed in U-Net 3+ [13] to alleviate this limitation, while leading to huge computational complexity. Besides, full-scale skip connections assume that all channels of the feature maps generated by different layers share equal weights. Nevertheless, features generated at different stages often possess different levels of discrimination. To tackle the inadequate utilization of features while reducing computational costs, we propose multi-scale skip connections with channel attention blocks to combine the multi-scale features and realign channel-wise features adaptively.

The importance of positions on the central skeleton of a square convolution kernel outweighs those on the corners [14, 15]. Thus, we utilize an asymmetric convolution block to enhance the representation capacity of convolution layers by strengthening the weight of the central crisscross parts. Asymmetric convolution blocks (ACB), involving branches of the square, horizontal and vertical kernels, could capture refined features by adding three convolution outputs together, without an increase in computational complexity [14]. The effectiveness of ACB has been demonstrated in other domains, such as image classification [14], image demoireing [16, 17], and medical image segmentation [18]. Here, we incorporate

*These authors contributed equally to this work.
This work was supported in part by the National Natural Science Foundation of China (No. 41671452). *(Corresponding author: Rui Li.)*
R. Li and S. Zheng are with School of Remote Sensing and Information Engineering, Wuhan University, Wuhan 430079, China (e-mail: lironui@whu.edu.cn; syzheng@whu.edu.cn).
C. Duan is with the State Key Laboratory of Information Engineering in Surveying, Mapping, and Remote Sensing, Wuhan University, Wuhan 430079, China; chenxiduan@whu.edu.cn (e-mail: chenxiduan@whu.edu.cn).
C. Zhang is with Lancaster Environment Centre, Lancaster University, Lancaster LA1 4YQ, United Kingdom; UK Centre for Ecology & Hydrology, Library Avenue, Lancaster, LA1 4AP, United Kingdom (e-mail: c.zhang9@lancaster.ac.uk).
P.M. Atkinson is with Lancaster Environment Centre, Lancaster University, Lancaster LA1 4YQ, United Kingdom; Geography and Environmental Science, University of Southampton, Highfield, Southampton SO17 1BJ, United Kingdom; and Institute of Geographic Sciences and Natural Resources Research, Chinese Academy of Sciences, 11A Datun Road, Beijing 100101, China (e-mail: pma@lancaster.ac.uk)

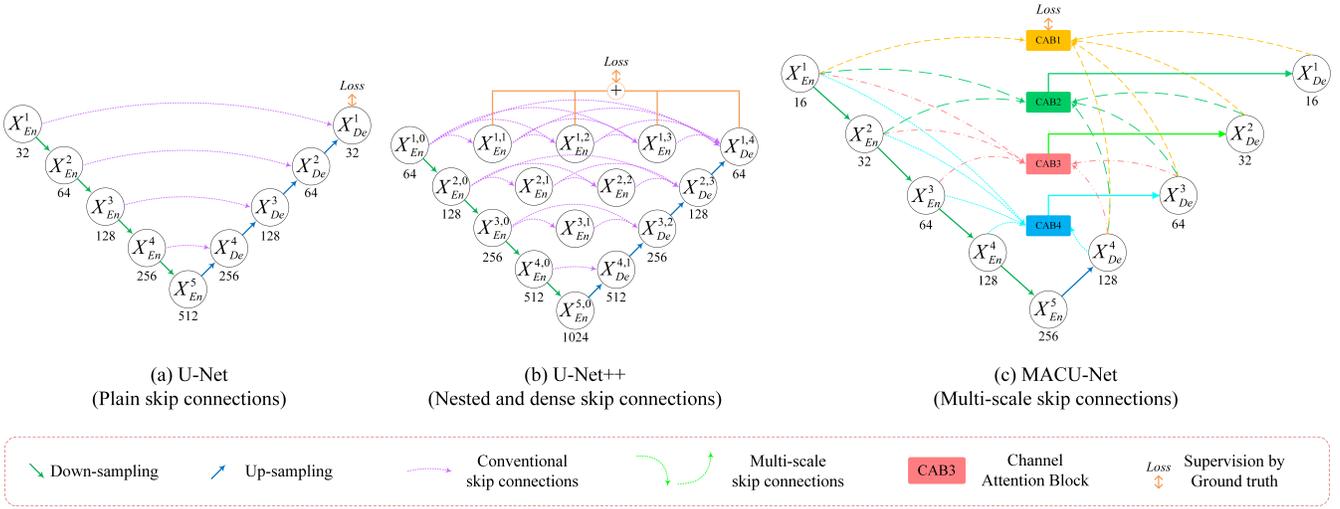

Fig. 1. Comparison of (a) U-Net, (b) U-Net++, and proposed (c) MACU-Net 3+. The depth of each node is presented below the circle.

ACB into U-Net for remotely sensed semantic segmentation by enhancing the representation capability of a standard square-kernel layer.

Based on the insights above, we design a multi-scale connected and asymmetric-convolution-based U-Net (MACU-Net) with asymmetric convolution blocks. To test the effectiveness of MACU-Net, we compare the performance of the proposed method with U-Net [11], FGC [19], U-Net++ [12], U-NetPPL [20], WRAU-Net [21], CE-Net [22] and U-Net 3+ [13]. The major contributions of this Letter are listed as:

1) We design multi-scale skip connections with channel attention blocks to make use of multi-scale features and realign channel-wise features.
2) We utilize asymmetric convolution blocks to substitute the standard convolutional layers, which enhance the representation capability of the convolution layers.
3) Based on multi-scale skip connections and asymmetric convolution blocks, we design a novel MACU-Net method and compare it with benchmark approaches comprehensively.

## II. METHODOLOGY

Fig.1 provides a graphical overview of U-Net, U-Net++, and the proposed MACU-Net. In comparison with U-Net and U-Net++, the multi-scale features of MACU-Net are integrated by the re-designed skip connections.

### A. Asymmetric Convolution Block

As reported in [14], a square convolution kernel captures features with uneven proportions. More specifically, the weights on the central crisscross positions (i.e., the skeleton of a kernel) have greater magnitude, whereas the points at the corners contribute less information to the feature extraction. Thereby, the cross-like receptive field can mitigate the influence of redundant information in capturing representative features, as illustrated in Fig. 2 (a).

We modify the asymmetric convolutions proposed in [14] and design an asymmetric convolution block (ACB) to capture features from different receptive fields as shown in Fig. 2 (b). There are three branches in ACB (i.e., a 3×3 convolution, a 1×3

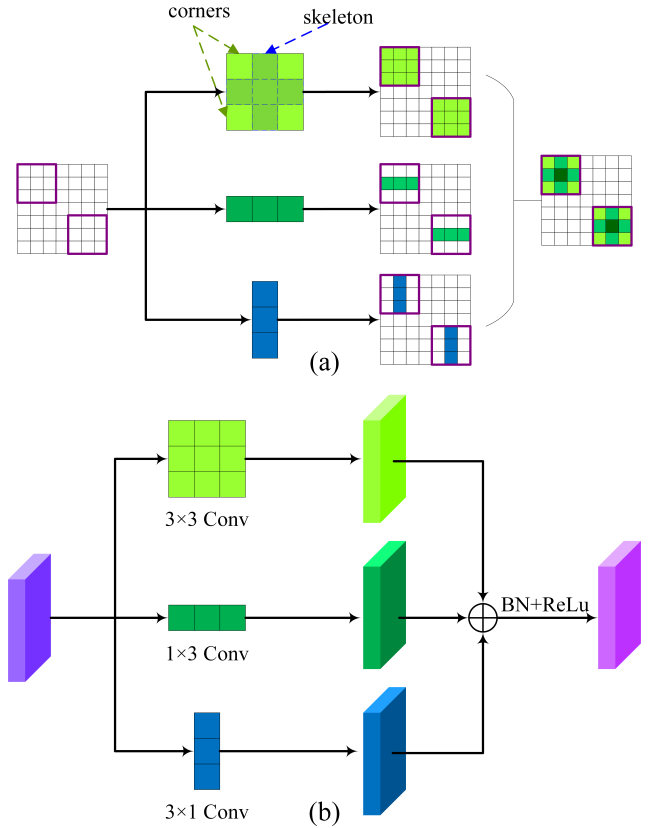

Fig. 2. (a) The ACB enhances the significance of the skeletons feature. (b) The structure of the asymmetric convolution block.

convolution (horizontal kernel), and a 3×1 convolution (vertical kernel)), to obtain a cross-like receptive field. The 3×3 convolution captures features by a relatively large receptive field, while the horizontal and vertical kernels guarantee the significance of features on the skeleton and expand the width of the network. The feature maps generated by the three branches are added to achieve the fusion results. Thereafter, batch norm (BN) and ReLU are used to increase the numerical stability and activate the output in a nonlinear manner. The formulation of ACB can be described as:

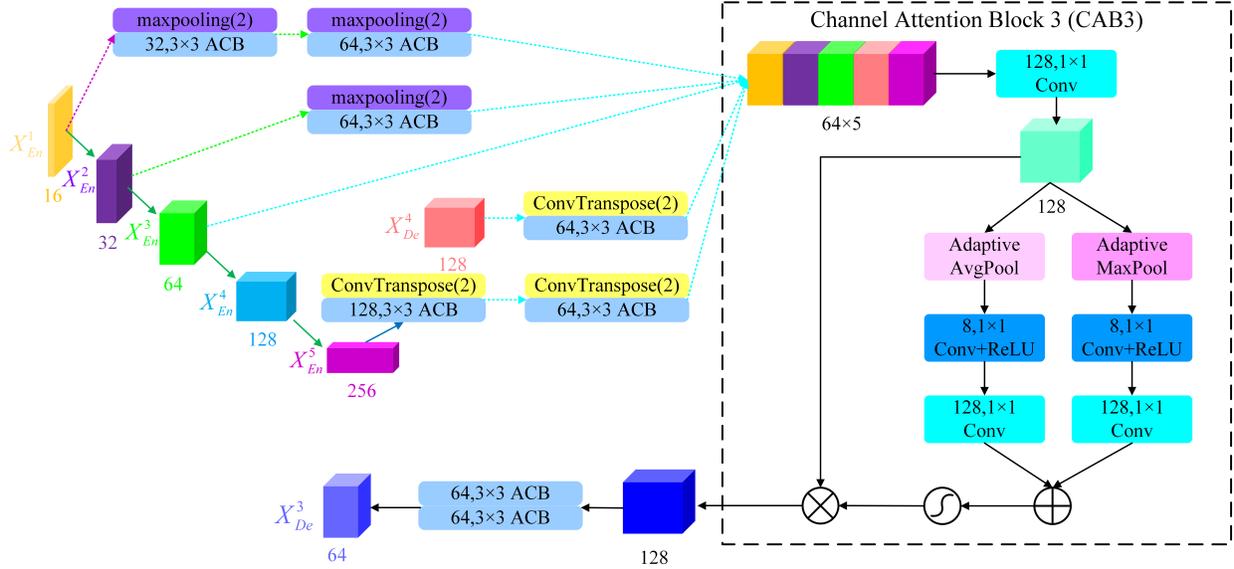

Fig. 3. Illustration of how to construct the multi-scale aggregated feature map of $X_{De}^3$.

$$\bar{x}_i = F_{3\times3}(x_{i-1}) + F_{1\times3}(x_{i-1}) + F_{3\times1}(x_{i-1}) \quad (1)$$

$$x_i = \sigma\left(\gamma_i \frac{\bar{x}_i - E(\bar{x}_i)}{\sqrt{Var(\bar{x}_i) + \epsilon_i}} + \beta_i\right) \quad (2)$$

where $x_i$ is the output of the ACB, and $x_{i-1}$ is the input of the ACB. $Var(\cdot)$ and $E(\cdot)$ represent the variance function and expectation of the input. $\epsilon$ is a small constant to maintain numerical stability. $\gamma$ and $\beta$ are two trainable parameters of the BN layer, and the normalized result can be scaled by $\gamma$ and shifted by $\beta$. $\sigma(\cdot)$ denotes the activation function of ReLU.

ACB is used to capture and refine the features in each layer of the encoder and is attached after each transposed convolution of the decoder, to avoid the checkerboard pattern and generate a smooth image.

### B. Multi-Scale Skip Connections

Since the information at multiple scales is not fully exploited by the plain connections of U-Net, we design multi-scale skip connections to capture the interplay between the encoder and decoder, which extract both fine-grained detailed information and coarse-grained semantic information.

Taking $X_{De}^3$ as an example, Fig. 3 demonstrates how to generate feature maps. First, the feature maps of the same-level encoder layer (i.e., $X_{En}^3$) are connected directly; Second, the fine-grained detailed information contained in lower-level encoder layers (i.e. $X_{De}^4$ and $X_{De}^5$) are delivered by transposed convolutions and asymmetric convolution blocks; Third, the coarse-grained semantic information contained in higher-level encoder layers (i.e. $X_{En}^1$ and $X_{En}^2$) are transmitted by the max-pooling layers and asymmetric convolution blocks. The above procedure can be formulated as:

$$X_{De}^i = \begin{cases} X_{En}^i, i = N \\ CAB\left(\left[\underbrace{A(D(X_{En}^k))_{k=1}^{i-1}}_{Scales:1^{th}\sim i^{th}}, \underbrace{A(U(X_{De}^k))_{k=i+1}^{N}}_{Scales:(i+1)^{th}\sim N^{th}}\right]\right), \\ i = 1, \ldots, N-1 \end{cases} \quad (3)$$

where $CAB$ indicates channel attention block which realigns channel-wise features, and $A(\cdot)$ denotes asymmetric convolution block. $D(\cdot)$ and $U(\cdot)$ represent down-sampling using max-pooling layers and up-sampling using transposed convolution, respectively, while $[\cdot]$ represents the operation of concatenation. Note the CAB4, CAB2, and CAB1 (as shown in Fig. 1. (c)) are connected similarly.

### C. Channel Attention Block

With five feature maps of equal size and resolution, we need to decrease the enormous number of channels further and realign channel-wise features. Motivated by the Convolutional Block Attention Module (CBAM) [23], we design the channel attention block (CAB) to reweight the channel-wise features as shown in the right of Fig. 3. The CAB aims to learn a 1-D weight $W_c \in R^{C\times1\times1}$ which realigns the channels of the input feature map $F \in R^{C\times H\times W}$, where $C$, $W$, and $H$ indicate the number of channels, height, and width of the feature map, accordingly. By multiplying $W_c$ and $F$, CAB enhances the informative channels and restrains the indiscriminative channels.

Taking $X_{De}^3$ as an example, we utilize a 1×1 convolution with 128 filters to reduce the number of channels initially. Thereafter, the spatial dimension is squeezed by the operation of an average-pooling and max-pooling simultaneously. By two convolution layers with eight filters and ReLU activation functions, the channels of the squeezed feature maps are compressed to one-sixteenth of their original size. The number of channels is then reinstated using two convolution layers with 128 filters. Finally, the sum of the two layers is activated by the sigmoid function and multiplied by the first convolution's output. Similarly, the $X_{De}^4$, $X_{De}^2$, and $X_{De}^1$ are generated by the corresponding CABs.

## III. EXPERIMENTAL RESULTS

This section first introduces the datasets and experimental settings to test the effectiveness of MACU-Net and then

TABLE I
THE EXPERIMENTAL RESULTS ON WHDLD (THE LEFT) AND GID (THE RIGHT) DATASETS.

| Method | OA | AA | K | mIoU | FWIoU | F1 | Method | OA | AA | K | mIoU | FWIoU | F1 |
| --- | --- | --- | --- | --- | --- | --- | --- | --- | --- | --- | --- | --- | --- |
| U-Net | 82.692 | 71.224 | 75.491 | 58.913 | 72.711 | 72.055 | U-Net | 82.257 | 84.399 | 77.485 | 73.226 | 70.472 | 84.069 |
| FGC | 83.721 | 69.866 | 76.571 | 59.465 | 73.524 | 72.479 | FGC | 82.827 | 84.394 | 78.162 | 74.094 | 71.603 | 84.728 |
| CE-Net | 83.982 | 71.768 | 77.065 | 60.878 | 73.982 | 73.604 | CE-Net | 83.378 | 85.747 | 78.975 | 74.487 | 72.111 | 84.942 |
| WRAU-Net | 83.989 | 72.346 | 77.106 | 61.253 | 74.006 | 74.143 | WRAU-Net | 83.083 | 85.386 | 78.533 | 74.707 | 71.730 | 85.099 |
| U-NetPPL | 83.802 | 72.300 | 76.886 | 60.787 | 73.836 | 73.751 | U-NetPPL | 83.599 | 86.682 | 79.271 | 75.207 | 72.449 | 85.419 |
| U-Net++ | 84.152 | 72.329 | 77.297 | 61.385 | 74.248 | 74.221 | U-Net++ | 84.050 | 85.880 | 79.777 | 75.508 | 73.135 | 85.657 |
| U-Net3+ | 84.094 | 72.630 | 77.317 | 61.256 | 74.314 | 74.101 | U-Net3+ | 84.271 | 84.847 | 79.983 | 75.203 | 73.545 | 85.463 |
| **MACU-Net** | **84.717** | **74.516** | **78.381** | **62.675** | **75.381** | **75.324** | **MACU-Net** | **84.719** | **87.133** | **80.654** | **76.414** | **73.973** | **86.221** |

compares its performance with those of different benchmark frameworks comprehensively.

### A. Datasets

The effectiveness of MACU-Net was tested using Wuhan Dense Labeling Dataset (WHDLD) [24, 25] and Gaofen Image Dataset (GID) [30]. WHDLD contains 4940 RGB images of 256 × 256 pixels captured by the Gaofen 1 and ZY-3 satellite sensors over the urban area of Wuhan, China. By image fusion and resampling, the attained spatial resolution of the images is 2 m. The images contained in WHDLD are labeled with six classes (i.e., bare soil, building, pavement, vegetation, road, and water). GID contains 150 RGB images of 7200 × 6800 pixels captured by the Gaofen 2 satellite sensor. Each image covers a geographic region of 506 $km^2$. The images contained in GID are labeled with six classes (i.e., built-up, forest, farmland, meadow, water, and others). We select 15 images contained in GID, which cover the whole six categories. The serial number of the chosen images is released along with our open-source code on GitHub.

### B. Experimental Setting

To evaluate the effectiveness of MACU-Net, the U-Net [11], FGC [19], U-Net++ [12], U-NetPPL [20], WRAU-Net [21], CE-Net [22] and U-Net 3+ [13] methods were used as benchmark comparators. Excluding U-Net, the remaining methods were adaptations of the raw U-Net.

All models were implemented with PyTorch and the optimizer was set as Adam with a 0.0003 learning rate, which decays according to cosine annealing. All the experiments were implemented on a single NVIDIA GeForce RTX 2080ti GPU. The cross-entropy loss function was used as the loss function.

For WHDLD, we randomly selected 60% of the images as the training set, 20% images as the validation set, and the remaining 20% images as the test set. For GID, we separately partitioned each image into non-overlapping patch sets of size 256 × 256 pixels and discarded pixels on the edges. Thus, 10920 patches were obtained. We selected randomly 60% patches as the training set, 20% patches as the validation set, and the remaining 20% patches as the test set.

For each dataset, the overall accuracy (OA), average accuracy (AA), Kappa coefficient (K), mean Intersection over Union (mIoU), Frequency Weighted Intersection over Union (FWIoU), and F1-score (F1) were adopted as evaluation indices.

### C. Results on WHDLD and GID

The experimental results of the different methods on WHDLD and GID are demonstrated in Table I. The performance of the proposed MAC-UNet transcends other algorithms in all quantitative evaluation indices. For WHDLD, the proposed MACU-Net brings a near 0.8% increase in mIoU compared with U-Net++. For the GID dataset, the gains are more than 0.9% in mIoU and 0.5% in the F1-score, respectively. Our method and U-Net's visual results are provided in Fig. 4, which demonstrate that the proposed MACU-Net can capture refined features. For example, the multi-scale skip connections and the asymmetric convolution enable the MACU-Net to discriminate the complicated boundary, which can be visibly seen in the final column in Fig. 4.

TABLE II
THE TRAIN TIME, COMPARISON OF PARAMETERS AND COMPUTATIONAL COMPLEXITY ON WHU DATASET.

| Method | Training time/epoch (s) | Parameters (M) |
| --- | --- | --- |
| U-Net | 239 | 10.858 |
| FGC | 213 | **2.189** |
| CE-Net | **161** | 29.005 |
| WRAU-Net | 299 | 6.36 |
| UNet-PPL | 307 | 32.587 |
| U-Net++ | 501 | 9.047 |
| U-Net3+ | 617 | 26.986 |
| MACU-Net | 298 | 5.152 |

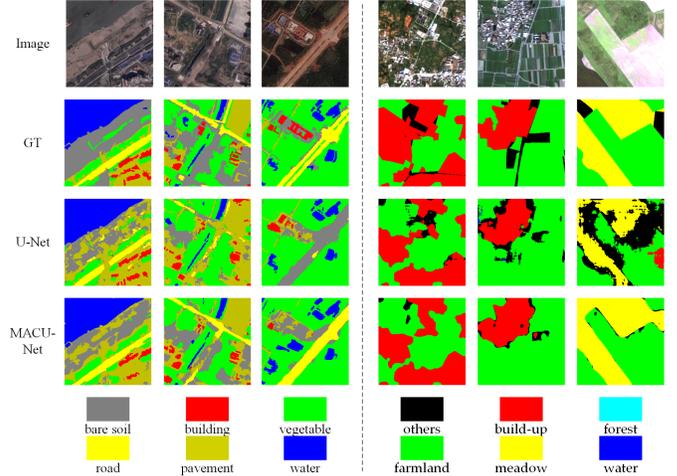

Fig. 4. Visualization of results on the WHDLD dataset (the left) and the GID dataset (the right).

The number of parameters and computational complexity was evaluated to assess quantitatively the framework efficiency. The comparison of the training time and parameters between different algorithms are reported in Table II, where 'M' represents million. The comparison demonstrates that the MACU-Net is space-efficient, with the second smallest number of parameters amongst the eight models. As for training time, even if the MACU-Net occupies less memory, MACU-Net is not faster than the raw U-Net but is significantly faster than U-Net++ and U-Net 3+.

*D. Ablation Study*

TABLE III
THE ABLATION STUDIES ON WHDLD (THE TOP) AND GID (THE BOTTOM).

| Method | OA | AA | K | mIoU | FWIoU | F1 |
|---|---|---|---|---|---|---|
| U-Net | 82.692 | 71.224 | 75.491 | 58.913 | 72.711 | 72.055 |
| U-NetH | 83.582 | 71.181 | 76.323 | 60.248 | 73.176 | 73.262 |
| U-NetV | 83.688 | 73.651 | 76.742 | 60.982 | 73.451 | 73.934 |
| ACU-Net | 84.098 | 72.699 | 77.325 | 61.436 | 74.283 | 74.267 |
| MU-Net | 83.573 | 72.772 | 76.708 | 60.738 | 73.730 | 73.678 |
| MACU-Net | 84.717 | 74.516 | 78.381 | 62.675 | 75.381 | 75.324 |
| U-Net | 82.257 | 84.399 | 77.485 | 73.226 | 70.472 | 84.069 |
| U-NetH | 83.164 | 84.470 | 78.631 | 74.063 | 71.908 | 84.693 |
| U-NetV | 83.327 | 84.582 | 78.804 | 74.033 | 72.165 | 84.692 |
| ACU-Net | 83.816 | 85.057 | 79.422 | 75.043 | 72.829 | 85.341 |
| MU-Net | 83.599 | 86.682 | 79.271 | 75.207 | 72.449 | 85.419 |
| MACU-Net | 84.719 | 87.133 | 80.654 | 76.414 | 73.973 | 86.221 |

To verify the impact of multi-scale skip connections and asymmetric convolution blocks, we analyzed the performance of the U-Net with the horizontal kernel (U-NetH), U-Net with the vertical kernel (U-NetV), U-Net with the ACB (U-NetHV), U-Net with multi-scale connections (MU-Net), and the proposed MACU-Net. The results shown in Table III demonstrate the effectiveness of multi-scale skip connections and asymmetric convolution blocks. Specifically, the utilization of the single horizontal kernel or the single vertical kernel will result in about 1% improvement of the OA, while the best practice is the joint use of both. Even if even the effectiveness is not as obvious as the ACB, the improvement of accuracy can also attribute to the multi-scale skip connections.

## IV. CONCLUSION

In this Letter, we design a multi-scale skip connected architecture, MACU-Net, for fine-resolution remotely sensed semantic segmentation. Based on multi-scale skip connections and channel attention blocks, semantic features generated by U-Net multi-level layers are combined and refined. Meanwhile, the representation power of the standard convolution layer is enhanced by the asymmetric convolution block. Numerical experiments conducted on two large-scale datasets confirm comprehensively the superior performance of our MACU-Net over benchmark approaches.